\documentclass{elsart}
\usepackage{graphicx}
\usepackage{dcolumn}
\usepackage{bm}

\begin{document}
\begin{frontmatter}
\title{Low-energy dipole excitations towards the proton drip-line: doubly magic $^{48}$Ni}
\author{N.~Paar\corauthref{cor},}
\author{P.~Papakonstantinou,}
\author{V.Yu.~Ponomarev\thanksref{home},}
\author{and J.~Wambach}
\corauth[cor]{Corresponding author {\em Email: }{\tt nils.paar@physik.tu-darmstadt.de}}

%%           ------INSTITUTIONS ---
%%
\address{Institut f\"ur Kernphysik, Technische Universit\"at Darmstadt,
Schlossgartenstrasse 9, D-64289 Darmstadt, Germany
}
\thanks[home]{Permanent address:
Joint Institute of Nuclear Research, Dubna, Russia}

%\author{N. Paar, P. Papakonstantinou, V. Yu. Ponomarev, and J. Wambach}
%\address{Institut f\" ur Kernphysik, Technische Universit\" at Darmstadt, Schlossgartenstrasse 9,
%D-64289 Darmstadt, Germany}
\date{\today}

\begin{abstract}
The properties of the low-energy dipole response are investigated 
for the proton-rich doubly magic nucleus $^{48}$Ni,
in a comparative study of two microscopic models:
fully self-consistent Relativistic Random-Phase Approximation(RRPA)
based on the novel density-dependent meson-exchange interactions, and
Continuum Random-Phase Approximation(CRPA) using
Skyrme-type interactions with the continuum properly included.
Both models predict 
the existence of the low-energy soft mode, i.e. the proton pygmy 
dipole resonance (PDR), for which the transition
densities and RPA amplitudes indicate the dynamics of loosely bound
protons vibrating against the rest of the nucleons.
The CRPA analysis indicates that the escape width for the proton
PDR is rather large, as a result of the coupling to the continuum.  

\end{abstract}
\end{frontmatter}

One of the major challenges in the region of unstable nuclei is the
understanding of soft modes of excitations which involve loosely
bound nucleons, nucleon halo,
or skin. In particular, in neutron-rich nuclei, nucleons
from the neutron skin may give rise to a low-energy dipole 
mode, known as a pygmy dipole resonance (PDR)~\cite{Suz.90}.
Some experimental evidence about the low-lying dipole
excitation phenomena in neutron-rich nuclei is available
from (i) electromagnetic excitations in
heavy-ion collisions ~\cite{Lei.01}, and (ii) nuclear resonance fluorescence
experiments for nuclei with moderate neutron to proton number
ratios, e.g.  $^{116,124}$Sn~\cite{Gov.98}, $^{138}$Ba~\cite{Her.99},
Pb~\cite{End.03}, and Ca~\cite{Har.04} isotopes, and N=82 isotones~\cite{Zil.02}. 
On the other side, a variety of theoretical models have in recent
years been employed
in studies of the low-lying E1 strength: continuum quasiparticle RPA formulated in the
coordinate-space Hartree-Fock-Bogoliubov framework~\cite{Mat.01},
continuum RPA with Woods-Saxon potential for the ground state 
and Landau-Migdal force in the residual interaction~\cite{Lei.04}, 
the time-dependent density-matrix theory~\cite{Toh.01}, 
Skyrme Hartree-Fock + quasiparticle  RPA with phonon
coupling~\cite{Col.01,Sar.04}, and
the quasiparticle-phonon model~\cite{Gov.98,End.03,TLS.04a}.
In the relativistic framework, properties of low-energy excitations 
have been systematically studied in the relativistic RPA (RRPA)~\cite{Vrepyg2.01},
and the relativistic quasiparticle RPA (RQRPA)~\cite{Paar.03,Ma.04,Paar.05,Paar_ijmp.05}.
Both experimental and theoretical studies qualitatively agree on the 
global properties of the isovector dipole response,
i.e., as the number of neutrons increases along the isotope chain,
the transition strength distribution is characterized by the appearance
of pronounced low-lying E1 strength.
Within the relativistic R(Q)RPA studies, it has been shown in a fully 
self-consistent way that the low-lying pygmy state represents a genuine
structure effect: the neutron skin oscillates against the core, exhibiting 
a collective nature from medium towards heavier nuclei~\cite{Vrepyg2.01,Paar.03}.
A detailed quantitative description of the low-lying E1 strength in neutron-rich
nuclei is essential for calculations of radiative neutron capture rates in the
r-process, and elemental abundances from nucleosynthesis~\cite{Gor.04}.

The structure of nuclei on the proton-rich side is equally important for
revealing many aspects of the underlying many-body problem and properties of
the effective nuclear interactions. At the same time, a quantitative description
of proton-rich nuclei is of a particular importance for describing the rapid proton
capture process of nucleosynthesis. Proton-rich nuclei are characterized
by unique ground-state properties such as $\beta$ decays with large $Q$ values, 
and direct emission of charged particles. The two-proton ground state radioactivity
has recently been observed as a decay mode of $^{45}$Fe~\cite{Gio.02,Pfu.02}.    
At the present time, knowledge on dipole excitations in nuclei
towards the proton drip-line is rather limited. Only very recently, the first microscopic 
theoretical prediction of the proton pygmy dipole resonance has been
indicated in the framework of RQRPA based on the Relativistic Hartree-Bogoliubov
model (RHB)~\cite{Paar_pp.05}. For nuclei close to the proton drip-line, the model
calculations predicted the occurrence  of pronounced dipole peaks below 10 MeV
in excitation energy, due to collective vibrations of loosely-bound protons against
the proton-neutron core. 

In this letter, we employ two theoretical approaches in a comparative study
of the low-lying E1 strength of the proton-rich nucleus \nuc{48}{Ni}:
the fully self-consistent Relativistic RPA (RRPA) based on novel
density-dependent interactions, and the Continuum RPA (CRPA) with Skyrme-type
interactions. An essential objective of this study is to ensure that
the proton PDR is inherent for different models, based on different
assumptions and effective interactions,
and to quantify the 
global properties of this mode, i.e. 
excitation energies and B(E1) strength, which may also be interesting for
the future experimental studies. Furthermore, we employ the CRPA model
to investigate the role of the coupling to the continuum for the proton PDR. 
We choose to analyse a doubly magic nucleus $^{48}$Ni, which is the most
proton-rich isotope that has been experimentally discovered~\cite{Bla.00}. 
For comparison, we also present results for \nuc{56}{Ni}, 
which lies close to  the valley of stability. 
Being doubly magic, these two isotopes can be studied within the CRPA and
RRPA models, which do not include pairing correlations.
Ground state properties of proton-rich nuclei around  $^{48}$Ni have
extensively been studied within the shell-model~\cite{Orm.96}, Hartree-Fock
Bogoliubov~\cite{Naz.96}, and Relativistic Hartree-Bogoliubov (RHB)
theory~\cite{Vre.98}. 
Proton drip-line nuclei are characterized by a 
reduction of the spin-orbit term of 
the effective interaction, outer orbits appear
to be very weakly bound, and the Fermi energy level may become positive. Due to the
presence of the Coulomb barrier, loosely bound orbits are
stabilised, in contrast to the neutron drip-line nuclei where the weakly bound
neutron orbits are more spatially extended. The interplay of all these effects
in the nuclear ground state will shape the properties of the corresponding 
dipole excitation response.

In the following, we briefly present the basic theoretical background of the 
Dirac-Hartree+RRPA and Skyrme-Hartree-Fock+CRPA models and
their effective interactions. 

In recent years, the relativistic mean-field theory and 
linear response based on density-dependent interactions, turned out to
be very successful in studies of nuclear ground-state properties and
excitation phenomena with a minimal set of parameters in a fully
microscopic way~\cite{Vre.05}.
Within this framework, the nucleus is described as a system of
Dirac nucleons which interact in  a relativistic covariant manner by
exchange of effective mesons. The model is formulated with the
Lagrangian density which explicitely includes the density dependence
in $\sigma$, $\omega$, and $\rho$ meson-nucleon vertices. 
In the present study, we employ the density-dependent effective
interactions which are constrained by the properties of finite
nuclei and nuclear matter: DD-ME1~\cite{Nik1.02},
and the new interaction DD-ME2
which provides an improved description of the isovector
dipole response~\cite{Lal.05}.
In the small-amplitude limit, the RRPA equations are derived from
the equation of motion for the nucleon density \cite{Nik2.02}. 
Both in RRPA and CRPA models, we use the same form of the 
effective isovector dipole
operator as in Ref. ~\cite{Paar_pp.05}.
The RRPA configuration space is constructed from particle-hole ($ph$)
pairs composed of the particle states above the Fermi level, and
hole states in the Fermi sea. In addition, in the relativistic
case, one also needs to include transitions to unoccupied states from the Dirac 
sea~\cite{Vre.05}. The resulting RRPA discrete spectra are averaged
with the Lorentzian distribution which includes an arbitrary choice for the
width, $\Gamma_{RRPA}$=1 MeV. 
The Dirac-Hartree+RRPA model is fully self-consistent, i.e. both the
equations of the ground state, and the residual RPA interaction are derived
from the same effective Lagrangian. This is an essential property for
an accurate decoupling of the spurious center-of-mass motion
without need for including any additional free parameters.
For the present study we use the Dirac-Hartree model formulated in the harmonic oscillator basis. Within this approach, the particle
continuum is represented by a set of discrete states, which are 
used to construct the RRPA configuration space. In Ref.~\cite{Vre.98}
it has been verified that for nuclei towards the proton drip-line, an
expansion in a large oscillator basis (N=20) provides sufficiently
accurate solutions, in complete agreement with the model formulated 
in the coordinate space.
 
The second theoretical framework for the present study is 
the Skyrme-Hartree-Fock (SHF) plus Continuum-RPA (CRPA) model. 
The HF equations describing the
ground state are derived variationally from the Skyrme energy functional. 
The $ph$ residual interaction is derived from the same energy functional. 
In the present study, the Coulomb interaction, as well as spin-dependent
terms are omitted from the residual interaction. 
The CRPA is formulated in the coordinate space,  
and the particle continuum is fully taken into account. 
The transition strength distribution $R(E)$ is continuous by construction. 
A small but finite value of 
Im$E\equiv\Gamma /2$ entering the evaluation of 
the $ph$ Green function 
ensures that bound transitions acquire 
a finite width and thereby contribute to the distribution. 
More details on the CRPA model can be found in Refs.~\cite{PWP2004,PWP2005} 
and references therein. 

For the purposes of the present study we implement various parameterisations of 
the Skyrme interaction, corresponding to different nuclear-matter properties. 
In particular, they have different (isoscalar) effective mass 
$m^{\ast}/m$ and isovector effective mass $m^{\ast}_v/m$. 
These quantities have influence on the evaluated properties of the 
isovector giant dipole resonance (IVGDR) \cite{Gor.04,Rei1999} and the 
low-energy dipole transitions. 
In general, interactions with a high effective mass are not able to 
reproduce the IVGDR properties. 
However, a high $m^{\ast}/m$ 
may be more appropriate to describe correctly the density
of states lying close to the Fermi energy, which are relevant for the present study. Therefore, we use several different parameterisations of the Skyrme
interaction, in order to ensure the general validity of our conclusions,
at least on a qualitative level.  
We employ parameterisation MSk7 ($m^*/m = m^{\ast}_v/m=1.05$), based on a Hartree-Fock-BCS model \cite{GTP2001},
and two interactions from the recent BSk series, namely BSk8 \cite{SGB2004} 
($m^{\ast}/m=0.80$, $m^{\ast}_v/m=0.87$) and 
BSk2 ($m^{\ast}/m=1.04$, $m^{\ast}_v/m=0.86$) \cite{Gor2002}. 
Both BSk interactions were parameterised by fitting the values 
of nuclear masses calculated within the Hartree-Fock-Bogoliubov method 
to essentially all the measured ones.  
We also use the traditional parameterisation SkM* ($m^*/m = 0.786$, $m^{\ast}_v/m=0.875$) \cite{Bar1982}, which was extensively used
in previous studies of giant resonances and response of exotic 
nuclei, e.g. Refs.~\cite{PWP2004,PWP2005,HSZ1997b,HSZ1998}. 

Next, we present the results obtained with the two models. 
In Fig.~1 we plot the ground-state proton and neutron density
distributions for $^{48}$Ni and $^{56}$Ni, calculated with 
Dirac-Hartree and SHF models, based on the DD-ME1 and
BSk8 effective interactions respectively. For  $^{56}$Ni
($Z=28$,$N=28$) the proton and neutron density distributions 
are similar in the nuclear interior and beyond the   
surface region the differences completely vanish. However,
in the case of  $^{48}$Ni ($Z=28$,$N=20$), due to the excess
of loosely-bound protons, the proton density distribution
is considerably extended
beyond the neutron density distribution. This effect is 
especially pronounced at radial distances $r>2$~fm, and it resembles
in structure a proton skin. Due to the presence of the 
Coulomb barrier, which tends to localise protons within the nuclear
interior, the proton skin is not so pronounced an effect as
the neutron skin. However, some evidence for increasing of the proton-skin
thickness in nuclei towards the proton drip-line is provided both by theoretical
and by experimental studies~\cite{Naz.96,Oza.02}. 

In Fig.~2 we present results obtained within the 
fully self-consistent RRPA model based
on Dirac-Hartree ground state with DD-ME1 effective interaction. 
In the left panel
we display the isovector dipole strength distributions for  $^{48}$Ni
and $^{56}$Ni. In the region of the isovector giant dipole resonance (IVGDR), 
the difference between the two distributions is very small and mainly consists
of small fluctuations of the IVGDR tail. In agreement with
the mass dependence of the giant resonance, the distribution of   $^{48}$Ni
is only slightly pushed to higher energies from the one for $^{56}$Ni.
However, in the low-energy region, the E1 strength distributions are rather
different: whereas there are no low-lying states for $^{56}$Ni, proton-rich
 $^{48}$Ni is characterized by the appearance of a pronounced amount of
low-lying transition strength. In order to clarify the origin of this
strength, in the right panel of
Fig. 2 we show the neutron and proton transition densities for two characteristic
peaks: the low-lying state at 7.72 MeV, and the giant resonance state
at 18.71 MeV. The transition densities of the latter display the dynamics
of IVGDR: collective oscillation of neutrons against protons.
For the low-energy peak, however, the proton and neutron transition
densities are in phase in the nuclear interior, whereas beyond the
surface region there are no contributions from the neutrons and proton
transition density dominates. This type of nuclear dynamics is characteristic
for the proton PDR, where loosely bound protons oscillate against the 
rest of the nucleons~\cite{Paar_pp.05}, in analogy to the neutron PDR in neutron-rich nuclei. 
The RRPA amplitudes for the E1 state
at 7.72 MeV reveal the structure of the proton PDR in detail: the main contribution
to the strength of the peak consists of transitions from  the proton 1f$_{7/2}$ state,
which is located at 0.11 MeV and weakly bound partly due to the presence of the
Coulomb barrier. 
The contributions from other transitions are at least an order of magnitude
smaller. Therefore, the appearance of the low-lying proton PDR strength is directly
related to the pronounced proton density distributions from Fig. 1. 
Protons from the same loosely-bound orbit contribute to the exotic nuclear
structure of the ground state, and to the excitation phenomena of the proton PDR.
The collectivity of the proton PDR peak considerably increases in
open shell-nuclei, due to the increased number of 
two-quasiparticle configurations composed from many states around
the Fermi surface which are, in that case, partially occupied~\cite{Paar_pp.05}.

The properties of the low-lying dipole transition strength are strongly sensitive
on the proton excess. In comparison of the cases of $^{46}$Fe (from
Ref.~\cite{Paar_pp.05}) and  $^{48}$Ni (DD-ME1 interaction is employed in both 
cases), one can see that the GDR peak energy
only weakly changes (0.2 MeV) with addition of two more protons. 
The peak energies of the proton PDR mode, however,
lowers from 9.4 MeV towards 7.7 MeV for $^{46}$Fe and $^{48}$Ni, respectively.
Obviously, the properties of the low-lying dipole transition strength are more
sensitive to the variations of the nucleon excess than GDR. This type of behaviour 
indicates the nature of the proton PDR mode: as the number of protons
increases, the oscillations of loosely-bound protons acquire lower frequencies
due to their weaker binding. 

By using the Skyrme Hartree-Fock+CRPA model with BSk8 effective
interaction, we repeat the same study of the isovector dipole transition
strength for  $^{48}$Ni and $^{56}$Ni isotopes. In the left panel of Fig.~3, we
notice that for the two Ni isotopes the strength distributions are not very 
different for E$>$10 MeV. In agreement with the RRPA results, the CRPA low-lying
transition strength for $^{48}$Ni is strongly enhanced in comparison with 
the $^{56}$Ni case. 
In the right panel of Fig. 3 we plot the transition densities for
a low energy peak at 9.72~MeV and the IVGDR at 20.28~MeV.
The displayed CRPA transition densities are in full agreement with the RRPA 
calculations, i.e. the high-energy peak corresponds to the collective IVGDR
where protons oscillate versus neutrons, while the low-lying transitions 
reveal the nature of the proton PDR. 

Within the CRPA model, we do not obtain only one pronounced low-energy
PDR peak for \nuc{48}{Ni}, 
but rather a smooth continuum, slightly structured around 9-10~MeV. 
One of the structures is the 9.72~MeV peak for which the transition density 
was plotted in Fig.~3. 
This continuum is not an artifact of the small smearing 
parameter used, $\Gamma=0.05$~MeV. 
The particle threshold energy is $E_{\rm th}=3.3$~MeV.  
We have evaluated the
proton- and neutron-transition densities corresponding to various
values of  excitation energy below $E=10$~MeV, and verified 
that they are characterized by transitions of a similar nature
as the 9.72~MeV peak. The above discussion remains valid   
qualitatively when other Skyrme interactions are used. More numerical 
results are presented below.   

In Table~1 we compare the global properties of the proton PDR
for $^{48}$Ni, obtained using the RRPA model with DD-ME1 and DD-ME2 effective
interactions, and the CRPA model with Skyrme interactions BSk8, BSk2, MSk7 and SkM*.
We have calculated the summed low-lying strength $m_{0}$, %=\sum{B(E1)}$,
the energy-weighted strength  $m_{1}$ %=\sum{E \cdot B(E1)}$, 
and the centroid energy $m_{1}/m_{0}$ for excitation energies below 10~MeV, 
where $m_k$ corresponds to the $k-$th moment of the strength distribution. 
Since the CRPA strength distribution is continuous in this region, 
the choice of the cut-off value of PDR region at 10 MeV is somewhat arbitrary
and the definition 
of the centroid energy becomes ambiguous. For this reason, the 
respective results are placed inside brackets. 
In addition, we list the relative amount of the low-lying strength $m_1$ with
respect to the classical Thomas-Reiche-Kuhn sum rule 
TRK$=14.9\cdot (NZ/A)  e^{2}fm^{2}$ MeV. 
The centroid energies of the proton PDR for $^{48}$Ni are obtained 
around 8 MeV. 
The two different
models are in fair agreement for various interactions,
exhausting for the proton PDR from $\approx$ 0.9$\%$ (BSk8) towards
$\approx$ 1.6$\%$ (DD-ME1, MSk7) of the classical
TRK sum rule. 

We found that that BSk8 and SkM* gave the lowest values of low-lying PDR strength. 
In general, the Skyrme-type interactions with the lower value of $m^{\ast}/m$,
result in weaker low-energy transition strength.  
The amount of the low-lying strength is directly related
to the energy $e_f$ of the least bound proton single-particle 
state $1f_{7/2}$ and the particle threshold energy $E_{\rm th}$.  
For $^{48}$Ni, these quantities are as follows: 
$e_f=-0.88$~MeV and $E_{\rm th}=4.2$~MeV for SkM*, 
$e_f=-0.03$~MeV and $E_{\rm th}=3.3$~MeV for BSk8, 
$e_f=0.26$~MeV and $E_{\rm th}=3.0$~MeV for BSk2, and  
$e_f=0.35$~MeV and $E_{\rm th}=3.0$~MeV for MSk7.
As the energy of  $1f_{7/2}$ proton state changes to higher values,
the corresponding low-lying E1 strength is more enhanced.
It is surprising that the Skyrme interaction whose 
effective-mass properties differ the most 
from the ones of the relativistic forces, namely MSk7, 
gives the best agreement with the RRPA results. 
The opposite holds for BSk8 and SkM*, whose properties differ the least from 
the ones of the relativistic forces. 
We notice, moreover, that the BSk2 corresponds to almost as large an effective mass $m^{\ast}/m$ 
as the MSk7, and practically the same $E_{\rm th}$, 
yet its predictions for the PDR are closer to the BSk8. 
Its isovector effective mass, though, is low and almost the same as BSk8. 
Within the present CRPA model, 
a Skyrme interaction cannot provide as much low-lying strength as the 
RRPA model, unless a high value of isovector effective mass is used.  
It is an interesting trend, but it cannot be established only by the
study of a single nucleus. 
Finally, there does not seem to exist a correlation between the degrees of 
quantitative agreement of the RRPA and CRPA models on the properties of the PDR 
on one hand, and the $e_f$ value on the other.   
The absolute difference between the value of $e_f$ %=0.35$~MeV 
corresponding to the Skyrme force MSk7 
(best agreement with RRPA on PDR strength) and the RRPA 
result, 0.11~MeV, is almost as large as in the case of the
SkM* (the worst agreement).  
  
In the last row of Table~1, we show an estimate of the PDR width, given by the
mean deviation of the strength distribution up to 10~MeV, 
$\sigma = \sqrt{(m_2/m_0)-(m_1/m_0)^2}$. Different Skyrme 
interactions result in similar widths, 
$\sigma \approx$ 1.4 MeV. On the other side, the width 
evaluated from the RRPA discrete strength distribution with
the DD-ME1 interaction with a smaller effective mass, results in
$\sigma =$ 0.77 MeV. In the RRPA case it provides only a measure
of the fragmentation of the low-lying strength, and therefore, it is
smaller in comparison with the CRPA width which includes the
contributions of the continuum.

In conclusion, we have studied the low-lying dipole response of a representative 
case of a proton drip-line nucleus, namely 
the doubly magic $^{48}$Ni.
We have employed two
different microscopic models, Dirac-Hartree+RRPA, and 
Skyrme-Hartree-Fock+CRPA, with various effective interactions. Within the
latter approach, a proper treatment of the particle
continuum is included, enabling us to study the relevance of the continuum 
for the low-lying strength.  
The comparison of E1 strength distributions for $^{48}$Ni and $^{56}$Ni,
and transition densities, show
that the low-lying dipole strength is a fundamental property of the 
proton-rich nuclei, and it corresponds to the proton pygmy dipole
resonance, where loosely bound protons vibrate against the approximately
isospin-saturated proton-neutron core. 
The coupling to the particle continuum results in an 
enhanced width of the proton PDR mode, estimated around 1.4 MeV. 
However, it remains unresolved why the best 
agreement on the proton PDR properties, between the relativistic and
nonrelativistic models, is obtained for DD-ME1 and MSk7 interactions
which have quite different properties. Whereas DD-ME1 represents an 
advanced density-dependent interaction appropriate for the studies of
both stable and exotic nuclei, MSk7 has large isoscalar and isovector
effective mass, and its properties were recently improved in BSk series.
A consistent comparison would necessitate a proper
inclusion of the continuum in the RRPA model, and on the other side, a fully 
self-consistent CRPA model without neglecting of terms in the
residual interaction. Nevertheless, we have shown that the two different
models agree fairly well on the global properties of the proton PDR, and
provide a clear theoretical picture for its underlying nature.  
We hope that the future experimental studies towards the proton drip-line
will provide evidence for this exotic mode.

\bigskip 
\leftline{\bf ACKNOWLEDGEMENTS}

This work has been supported by the Deutsche Forschungsgemeinschaft (DFG)
under contract SFB 634.

\bigskip \bigskip
%=========================================================================

\newpage

\begin{figure}
\includegraphics[scale=0.6,angle=0]{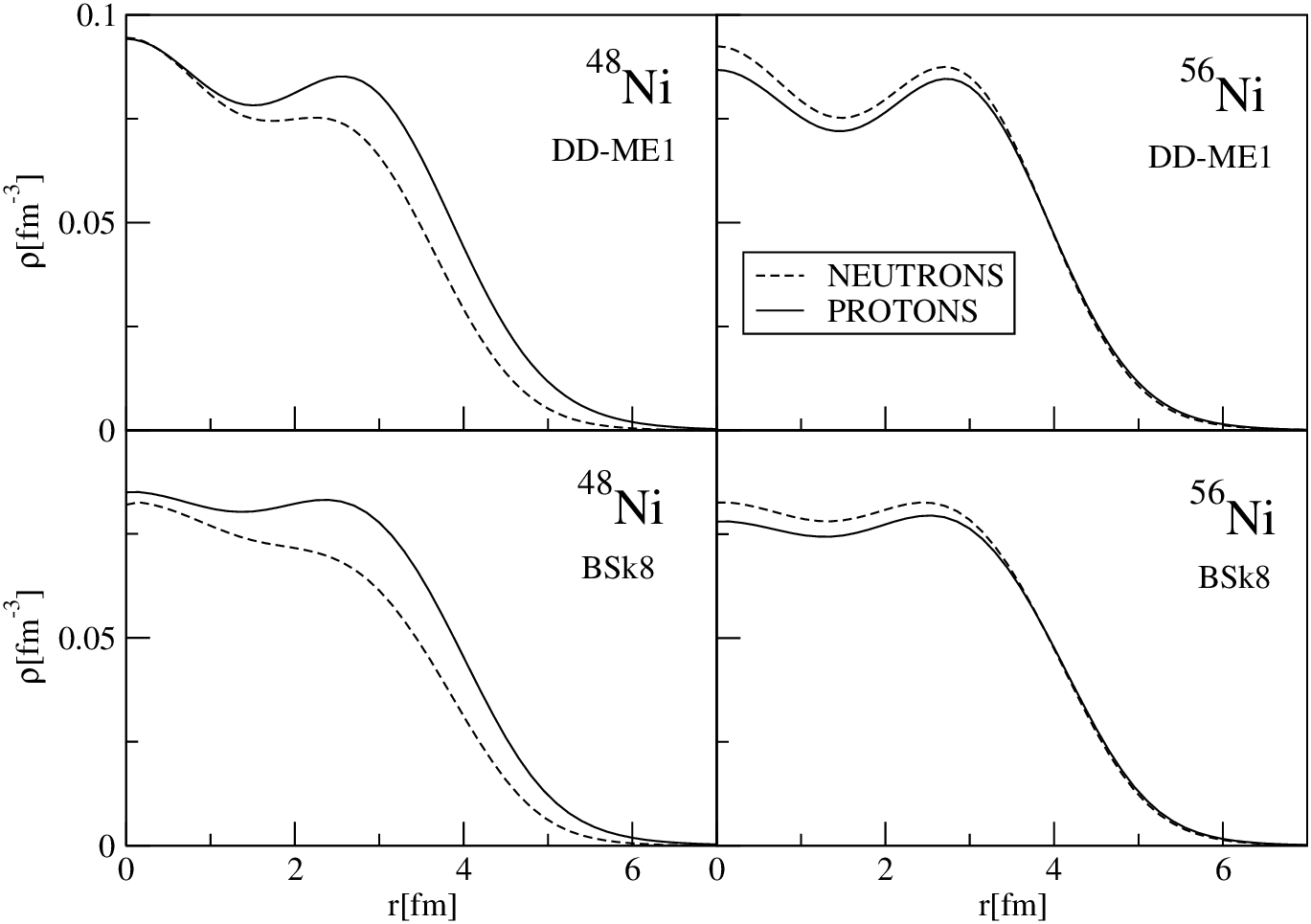}
\caption{Proton and neutron ground-state density distributions for $^{48}$Ni and 
$^{56}$Ni calculated with the Dirac-Hartree (DD-ME1 interaction) and
Skyrme-Hartree-Fock (BSk8 interaction) models.}
\label{fig1}
\end{figure}

\begin{figure}
\includegraphics[scale=0.6,angle=0]{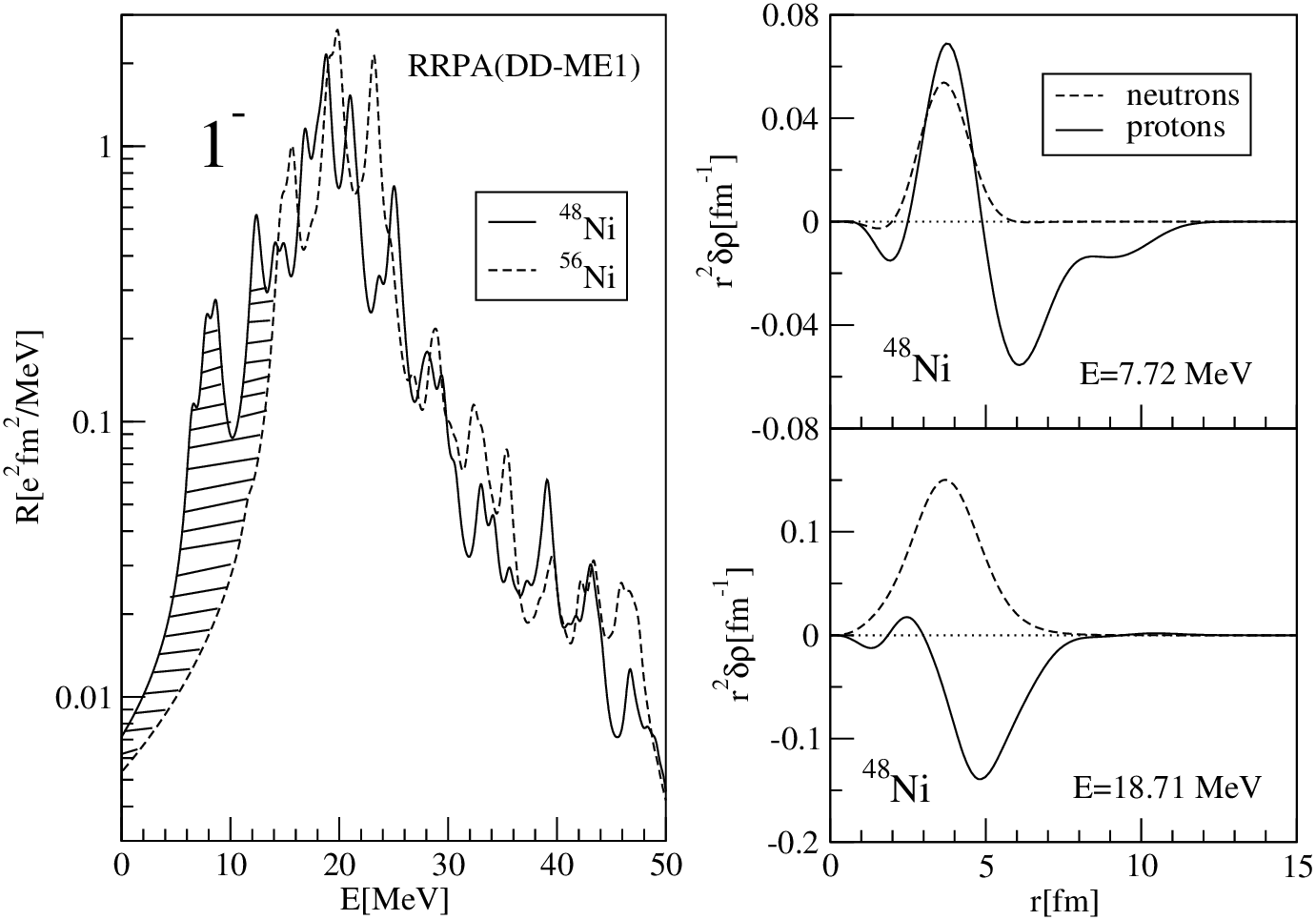}
\caption{The Dirac-Hartree+RRPA isovector dipole strength distribution 
for $^{48}$Ni and $^{56}$Ni (left panel). The proton and neutron
transition densities are displayed for the low-lying state (E=7.72 MeV) 
and isovector giant resonance (E=18.71 MeV). The DD-ME1 effective interaction
is employed. The shaded area denotes the difference between the low-energy
part of the strength distributions of $^{48}$Ni and $^{56}$Ni.}
\label{fig2}
\end{figure}

\begin{figure}
\includegraphics[scale=0.6,angle=0]{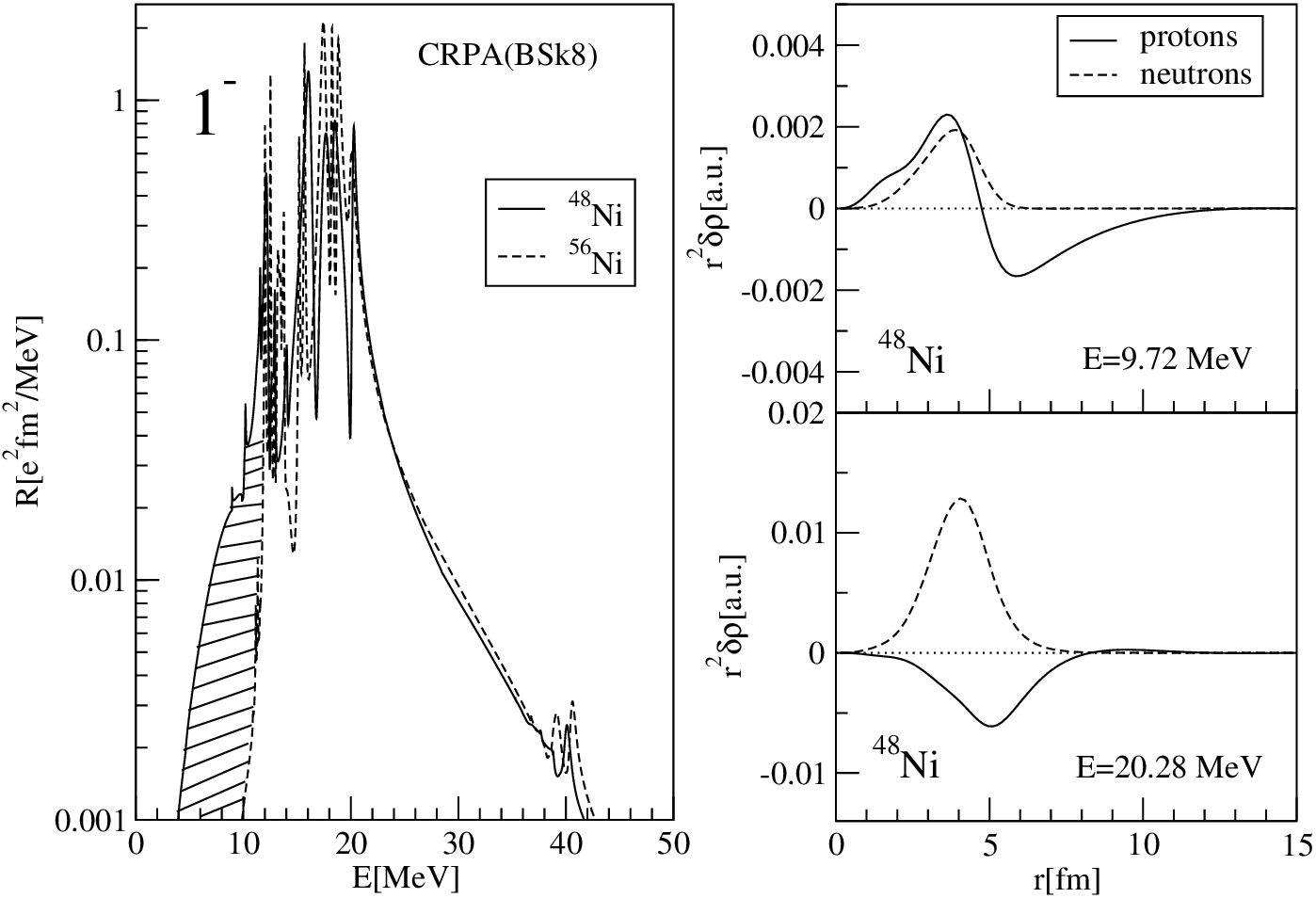}
\caption{Same as in Fig.~2, but calculated with the Skyrme Hartree-Fock+CRPA
model by using BSk8 effective interaction. Transition densities 
are shown for the states with E=9.72 and 20.28~MeV. 
}
\label{fig3}
\end{figure}

\begin{table}[htbp]
\centering
\begin{tabular}{ccccccc} \hline
                                  &  DD-ME1 & DD-ME2 & BSk8   & BSk2  & MSk7  & SkM* \\ \hline
$m_{0}[e^{2}fm^{2}]$              &  0.346  &  0.336 & 0.201  & 0.259 & 0.352 & 0.179  \\ \hline
$m_{1}[e^{2}fm^{2}MeV]$           &  2.783  & 2.718  & 1.632  & 2.132 & 2.914 & 1.485 \\ \hline
$m_{1}/TRK[\%]$                   &  1.60   &  1.56  & 0.94   & 1.23  & 1.67  & 0.85 \\ \hline
$m_{1}/m_{0}[MeV]$                &  8.03   &  8.08  & [8.10] & [8.22]& [8.28]  & [8.30]  \\ \hline
$\sigma [MeV]$                   &  0.77       &  0.79       & [1.42] & [1.45]& [1.48]  & [1.40]   \\ \hline
\end{tabular}
\caption{The low-lying (E$<$10 MeV) dipole transition strength ($m_{0}$), energy-weighted
low-lying strength ($m_{1}$), its fraction of the classical TRK sum rule,  
centroid energies and widths ($\sigma$) for $^{48}$Ni, calculated with RRPA (DD-ME1 and DD-ME2 effective interactions)
and CRPA (BSK8, BSK2, MSk7 and SkM* Skyrme effective interactions) models. 
The definition of the centroids and widths in the CRPA model is somewhat ambiguous (see text) and therefore the corresponding values are given in square brackets.  
}
\label{table}
\end{table}

\end{document}